# Patent-KG: Patent Knowledge Graph Use for Engineering Design

Haoyu Zuo[1], Yuan Yin[1], Peter Childs[1]

[1] *Engineering Design Group, Dyson school of Design Engineering, Imperial College London, South Kensington, London, SW7 2AZ*

## Abstract

To facilitate knowledge reuse in engineering design, several dataset approaches have been proposed and applied by designers. This paper builds a patent-based knowledge graph, patent-KG, to represent the knowledge facts in patents for engineering design. The arising patent-KG approach proposes a new unsupervised mechanism to extract knowledge facts in a patent, by searching the attention graph in language models. This method avoids using expensive labelled data in supervised learning or listing complex syntactic rules in rule-based extraction. The extracted entities are compared with other benchmarks in the criteria of recall rate. The result reaches the highest 0.9 recall rate in the standard list of mechanical engineering related technical terms, which means the highest coverage of engineering words. The extracted relationships are also compared with other benchmarks. The result shows that our method provides more contextual information in relationships, and extracts more relationship types including positional and negation relationships.

## 1. Introduction

The expansion in technological data every year is substantial. For example, in 2019, more than 3 million patents and 1.5 million scientific papers (World Intellectual Property Organization, 2019) were published worldwide. This can be contrasted with a human's reading capacity, 264 per papers per year (Van Noorden, 2014) on average. The ever-growing quantity of data provides both considerable challenges and potential advantages to researchers. On one hand, the fact is that it is hardly possible for an individual to fully search and comprehend a specific domain, as the published data is still growing explosively every day. On the other hand, Swanson (1986) hypothesized that a scientific discovery can be established by systematically studying the existing knowledge. The vast amount of data are of high diversity, and can be reused as incentives and stimuli for new knowledge. Reusing existing knowledge to speed up the idea generation has already been used in the domain of design.

To represent the knowledge facts in data, a new concept knowledge graph is introduced. A knowledge graph is defined (Wang et al., 2017) as "a multi-relational graph composed of entities and relations which are regarded as nodes and different types of edges, respectively" to represent the knowledge. A knowledge graph expresses knowledge in the format of a triple, which includes a head, a relationship and a tail. For example, consider Fig.1.1 illustrating the knowledge fact: Albert Einstein was born in German Empire. The knowledge triple is represented it as: (Albert Einstein, BornIn, German Empire) and it is transformed into the images with two nodes as the head and tail, one edge as the relationship. Many acknowledged facts can be composed together in a knowledge graph.

The aim of the paper is to construct a knowledge graph based on patent data, to represent the engineering knowledge facts and facilitate the knowledge reuse for engineering design.

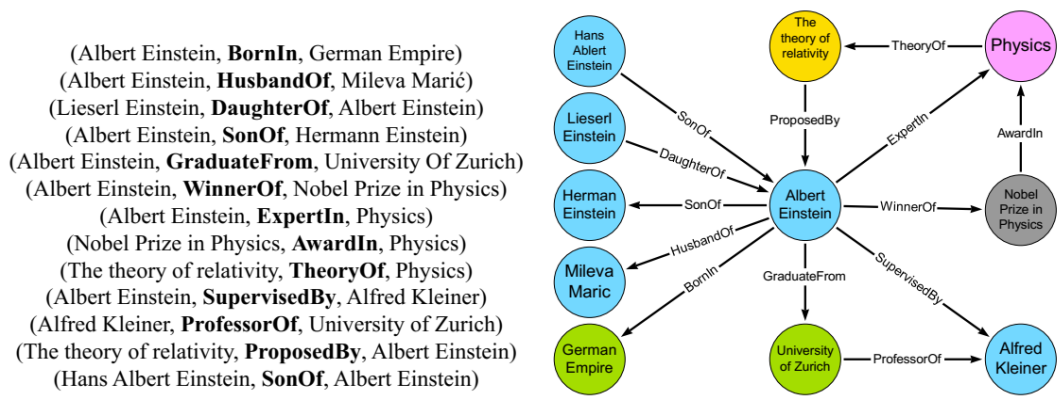

Fig.1.1 An example of a knowledge graph

## 2. Related work

### 2.1 Network for engineering design.

When considering knowledge reuse, common knowledge sources include encyclopedias, patent documents, scientific literatures, and reports. The growing need for engineering design drives researchers to construct a large database based on these sources. Shi (2018) proposed a pipeline to crawl design knowledge from design posts and the Elsevier published scientific literature, and constructed a structured ontology network to provide a semantic level understanding of the design knowledge. This semantic level representation of understanding were applied for design information retrieval and insight for idea generation process. Kim and Kim (2011)introduced a new way to represent the design knowledge in reports in a causal approach instead of the prevailing procedural approach. Both approaches were compared and a case study is conducted to demonstrate the causal approach is superior for reasoning capacity and knowledge cultivation. Sarica et al. (2020)trained a semantic engineering knowledge graph from the patent data, to overcome the limited coverage of traditional keyword retrieval. A case study demonstrated that the proposed method improves the efficiency to assist the early stage of design work. Chen et al. (2013)proposed an information system in Wikipedia to extract core information inside the selected articles. These pieces of extracted information were analysed in pairs and frequency to gain insight into relationships, and thus to support the conceptual design stage. Siddharth et al. (2021) uses the syntactic rules to extract knowledge triple (head, relationship, tail) from patent data to build an Engineering knowledge graph. Listing these rules is laborious and these rules can't be listed comprehensively, which will cause a partial extraction. Our goal is to extract the head, tail and relationship without these rules and achieves a better result. Patents were chosen as the data source as this form of documents records the content of inventions and contains a large quantity of scientific and technological information (Aristodemou and Tietze, 2018).

## 2.2 Information extraction

Information extraction is a task to identify and recognize words or phrases in text. This task is a fundamental activity in knowledge graph construction and various methods have been proposed. Methods can be divided into rule-based information extraction, statistic-based information extraction and neural network-based information extraction. (Li et al., 2020)

1) Rule-based information extraction

Rule-based information extraction (Li et al., 2020) extracts entities based on predefined rules. After analyzing the characteristics of entities, artificial rules of the intended information, such as syntactic rules or POS tagging, need to be constructed to match and identify the entity in the text. For example, to extract disease name such as type 1 diabetes, type 2 diabetes and type 3 diabetes, the rule is 'type + number + diabetes'. The limitation is obvious, the rule can never be listed and formulated comprehensively by human resources.

2) Machine learning based - supervised information extraction

The core method of machine learning is to simulate the human brain to build a neural network that can analyze and learn. Through the non-linear calculation of the hidden layer of the neural network, the features of words can be automatically combined and learnt. In recent ten years, neural network is experiencing explosive research and various models are proposed. These models are capable of extracting entities such as person name and place name(Luo et al., 2018, Huang et al., 2015). However, this method needs the tagging corpus, and the accuracy of recognition depends heavily on the tagging quality of the corpus.

3) Machine Learning- unsupervised information extraction

Supervised machine learning with labelled data is expensive for human labor, thus makes it can't be applied in a large-scale information extraction. So, weak supervised learning or unsupervised learning are proposed. For example, distant supervised learning (Mintz et al., 2009) will apply similar labelled data text to the target text, thus to avoid the tagging task. TextRunner(Etzioni et al., 2008) , Reverb(Fader et al., 2011) , and Ollie (Schmitz et al., 2012) are three unsupervised learning mechanism. These three methods transform knowledge facts in plain text into triples without predefined classification, which is also called open information extraction. However, the quality of unsupervised normally can't compare with supervised learning because noises will be introduced as no label data as filter. In Natural language processing areas, language models(LM) such as BERT and GPT-2(Radford et al., 2019) has achieved a great marks in some related tasks, such as: sentence classification(Wang et al., 2018). In the structure of LMs, it use a multi-head attentions that mimics cognitive attention, which will enhance the importance of the aiming part from the input data and fade the others. Wang et al. (2020) further applies the attention mechanism in language models for information extraction inside the sentence.

In summary, these traditional information extraction methods all have their own advantages and disadvantages. Patent data, for this study, are characterized by complexity of the language, some of the models developed in non-patent style are not applicable and need further modification. To overcome these difficulties, new methods are discussed and proposed. For example, Verberne et al. (2010) quantifies the main three challenges in dependency parsing for patents. Among those challenges, with the syntactic structure being the most problematic. Andersson et al. (2016) pointed out the difficulties in identifying the terms in patents, and proposed a new approach to combine linguistic methods and statistical method but without C-value. With this proposed approach, Andersson used 'termhood' to refer to 'The degree to which a stable lexical unit is related to some domain-specific concepts'. The termhood of the multi-word term were learned and reached the state of the art. Hu (2018) conducted a crowdsourced POS tagging patent task by Amazon Mechanical Turk, attempting to provide substantially more contexts while tagging.

# 3. Construction of patent KG

## 3.1 Data source

Patent data, for this study was gathered from 2016 to 2021, with Cooperative Patent Classification (CPC) codes start in 'F'(European Patent Office, 2021) – referring to "Mechanical engineering; lighting; heating; engines or pumps". Note that the patent data needs further filtering process to remove duplicate ones. In one patent family, there are normally more than one patent with different "Application ID" but with same name and almost same content. This problem is caused by publication variations in a later time, or a same version published in different country to protect their legal right internationally. The variations and other international publications are unnecessary for retrieving again and therefore are filtered and removed. By applying "Application ID" equals to "Application Earliest Filing ID", only the earliest patent in a patent family will be retrieved. In total, 457,815 patents were retrieved after filtering.

## 3.2 Data pre-processing

The purpose of this step is to process and prepare each sentence of the abstract into a lists of tokens.

First the abstract will be split into sentences with spaCy (Honnibal Matthew, 2017). Note that in spaCy, the default sentence segmentation will only split sentence on punctuation such as '.', '!' or '?' in a general language rule. In patent text, a period tends to be used in case of an abbreviation or to signify the end of one sentence or claim. Normally the sentence in a patent can be very long, the semicolon ';' is used frequently thus to allows one to incorporate multiple discrete parts of a sentence. Therefore the ';' token is added as a sentence boundary and overwritten in spaCy. In this way, the splitting can avoid long sentence and can be more accurate.

Second, the sentence will be split into tokens, and then noun phrases will be recognized and combined. Patents contain a large number of technical terms, with some of them are even new technical terms formed in the patent. There are three ways of technical terms formation in patents normally(Andersson et al., 2016):

(1) orthographical unit, e.g. bookcase, airplane, curveball. These words can be recognized by spaCy normally via its POS tagging.

(2) multi-word unit (MWU), e.g. airplane wings, knowledge graph, natural language processing. These words can be recognized by spaCy normally via its POS tagging.

(3) combined with hyphenation (e.g. H-theorem, mother-in-law). The default tokenizer in spaCy will split on hyphens. To avoid this, the existing infix definition is overwritten and a regular expression that treats a hyphen between letters as an infix is added,

Third, the sentence will be split into tokens, and then phrasal verbs will be recognized and combined. A phrasal verb is the combination of a verb and a paritcle, such as an adverb or a preposition, e.g. relate to, positioned through, engageable with. The verb contains the action information while the particle contains the additional information. Both can be part of the relationship, so it is also recognized and combined after tokenization. The sentence is parsed to find a token with dependency 'prt' (phrasal verb particle) and then the token's head with POS tagged 'VERB'.

### 3.3 Patent KG construction
#### 3.3.1 Dependency patterns

After the data pre-processing stage, the technical terms as nouns are recognized, the next step is to find the relationships (mostly verbs) between them. The patent texts are analyzed and the 5 interested dependency patterns(Marie-Catherine de Marneffe, 2016) are listed and corresponding examples are as follows:

(1) dobj: direct object

The direct object of a verb phrase is the noun phrase which is the (accusative) object of the verb.

**"The sensor sends a signal"     dobj(sends, signal)**

**"The electricity light the bulb"     dobj (light, bulb)**

(2) nsubj: nominal subject

A nominal subject is a noun phrase which is the syntactic subject of a clause. There are two scenarios, the governor of this relation is a verb or a copular verb. When the relation is a copular word, the root of the clause is the complement of the copular verb, which can be an adjective. However, the scenario of an adjective is not considered in patent patterns because it can't form a knowledge fact.

**"The baby is cute"     nsubj (cute, baby)  * not considered as knowledge facts***

**"She left him a note"     nsubj(left, she)**

(3) cop: coupla

A coupla is the relation of a function word used to link a subject to a nonverbal predicate, including the expression of identity predication.

**"HTR is the fourth generation nuclear power station"** cop (station, is)

**"Bill is a good person"** cop (person, is)

(4) neg: negation modifier

The negation modifier is the relation between a negation word and the word it modifies.

**"The temperature doesn't rise"** neg (rise, doesn't)

**"The method isn't working"** neg (working, isn't)

(5) pobj: object of a preposition

The object of a preposition is the head of a noun phrase following the preposition.

**"The colls in the axial magnetic bearings"** pobj (in, anxial magnetic bearing)

**"Place the card inside the slot"** pobj (inside, slot)

3.3.2 Match

The technical terms are grouped in two as a head and tail pair (h, t), then the match stage will find the best relationship between the (h, t) pair to generate a knowledge triple (h, r ,t) by searching the attention matrix. The attention matrix is formed within transformer-based language model BERT(Bidirectional Encoder Representations from Transformers).

BERT (Devlin et al., 2018)is a large and pre-trained Transformer network, with 12 layers where each layer consists of 12 attention heads. Fig 3.1 illustrates how the attention score is computed by selecting only one attention head in the first layer.

Firstly, the input tokens ($x_1,x_2,x_3,x_4$) are transformed into a sequence of vectors [$a_1,a_2,a_3,a_4$]. Then each vector is transformed into a query and a key vector by the linear transformation matrix $W_q$ and $W_k$. Starting with a query vector, e.g. $q_1$, the query vector will have a dot product with the key vector of all the other, e.g. $k_1$, $k_2$, $k_3$, $k_4$ (including the other key vectors and itself). Then softmax is applied over all the scores, $a_{1,1}$, $a_{1,2}$, $a_{1,3}$, $a_{1,4}$ to normalize them to be positive and sum to one.

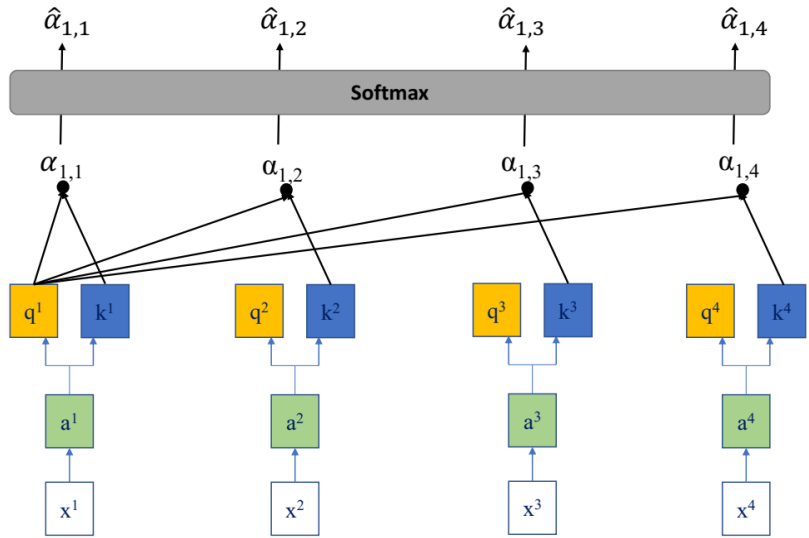

Fig.3.1 Computing process inside one head of attention

The attention mechanism will generate normalized weights $\hat{a}_{1,1}$, $\hat{a}_{1,2}$, $\hat{a}_{1,3}$, $\hat{a}_{1,4}$ which decide how "important" for each other token is when calculating the next representation on the current token. There are 12 heads in each layer, so more than one head enables BERT to learn more about the structure of the text. BERT also stacks multiple layers of attention, each of which computes based on the output of the previous layer. Through this repeated structure, the attention heads from deeper layer are able to form richer representation after previous computation. In total BERT's architecture comprises 12 layers with 12 heads, resulting in a total of 12×12 = 144 different attention heads. Clark et al. (2019) analysed and visualized all these attention heads, revealing that the attention heads include patterns such as finding direct objects of verbs, determiners of nouns, objects of prepositions, and objects of possessive pronouns. The research result also suggest that the certain head performs finding dobj at the accuracy of 86.8, nsubj at the accuracy of 58.5 and pobj at the accuracy of 76.3.

Fig 3.2 are 5 attention map examples on different interested attention patterns in patents, all these examples are computed on head 8-9. The line from one to another indicate the "importance" between each other, the deeper the color is, the more important it is. The aiming words in a sentence are colored red to highlight it from the others. For example, in the pattern of "dobj", the object word "signal" is colored red to see if the verb is allocated with higher weight. The line between "signal" and right verb "send" has the deepest color which means the attention head is recognizing the "dobj" pattern. Fig 3.3 are some negative examples from other attention heads, these example results are not in line with our interested patterns in patents but have some other attention patterns.

The heads in the same layer are found to be close to each other in the cluster, which indicates that these heads have similar pattern model toward some syntactic rule. Based on Clark's result, the 5 listed dependency patterns in patents are covered after the 9th layer. So, instead of using the attention output from the last layer of the encoder, the attention score is calculated in the BERT's architecture, and the attention is the mean of the 12 heads of the $9^{th}$ layer.

The attention mechanism in BERT calculates the attention from token to token. However, the words in our sentence are split into tokens and then re-combined if noun phrases and phrasal verbs are recognized. Therefore the attention is converted from a token-to-token map to word(phrase)-to-word(phrase) map. For attention from a phrase, the mean of the attention weights is calculated over the tokens. For attention to a phrase, the sum of attention weights is calculated over the tokens. These transformations preserve the property that the attention from one to other sums to be one.

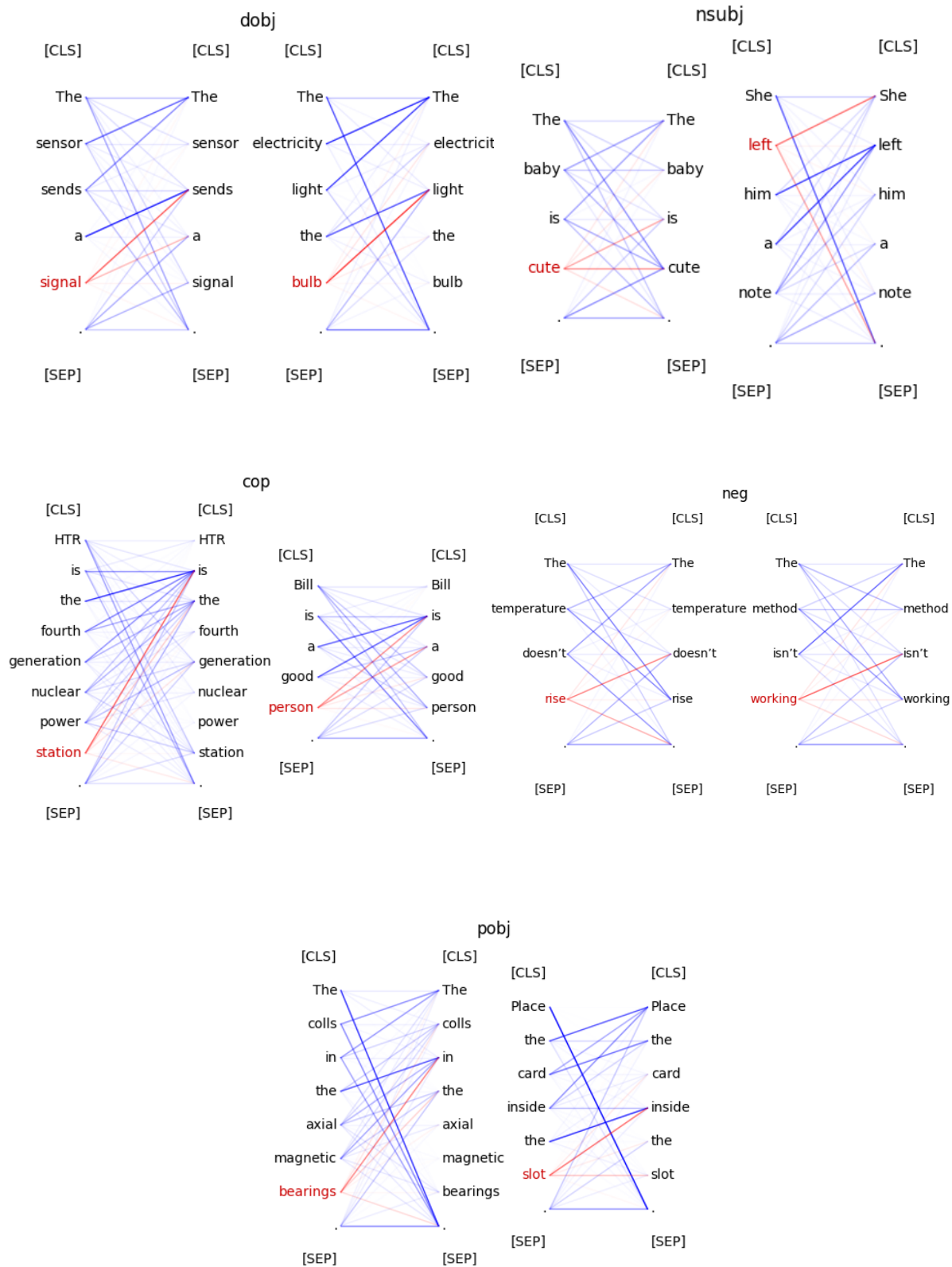

Fig.3.2 Attention examples on language patterns.

After the calculation, the attention graph(matrix) will be obtained such as showing in Fig.3.4 (on the right). To interpret this, the rows mean the key (from), the columns mean the query (to). In the attention graph, the beam search is used to find the best matched relationship (mostly verbs) candidate fact. For every head and tail pair (h, t) in a sentence, the beam search will search backwards: t→r→h, computing through the k words with k-highest attention scores between the head and tail. Taken the sentence "the magnetic force provided levitates the shaft" as example, the head-tail pair is (the magnetic force, the shaft), the beam size equals 2, the search computing processes are as the following:

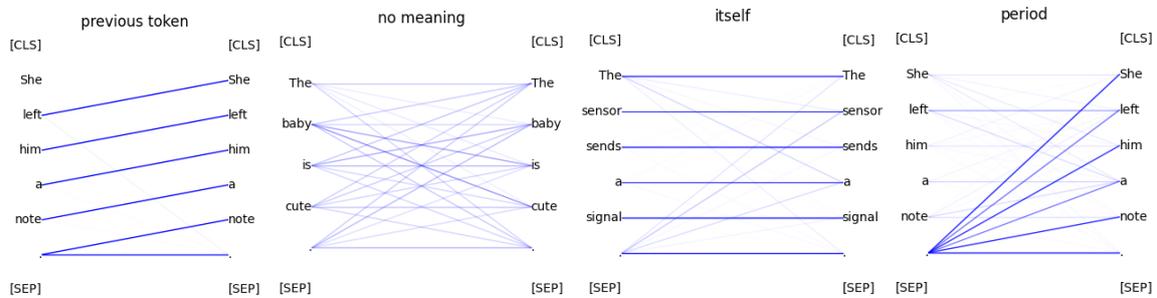

Fig.3.3 Negative attention examples.

(1) First, instead of searching forward, the searching algorithm is searching backward. The reason for searching backward is that, the later words has stored the knowledge from previous words, while the previous words may haven't read about later words. We need to add the tail "the shaft" to the beam, mark the head "the magnetic force" as the ending position and initialize the total attention degree as 0.

(2) Find the token with the largest attention score with the tail "the shaft", add that to the candidate (the shaft, levitates) and update the attention score 0.7761. Mark the token "levitates" as added to prevent search again. Then find the attention score between relationship "levitates" and "the magnetic force" which is 0.2496. The total attention score is 0.7761+0.2496= 1.0257

(3) Find the token with the second largest attention score with the tail "the shaft", add that as a candidate, (the shaft, provided) and update the attention score 0.0154. Mark the token "provided" as added to prevent search again. Then find the attention score between relationship "provided" and "the magnetic force" which is 0.5684. The total attention score is 0.0154+0.5684=0.5838

(4) The search will be stopped because the number of candidates reached the limit of beam size 2, and also reached the marked ending position "the shaft". The two candidate facts now both have attention scores from tail to relationship and relationship to head. The candidate with the highest attention scores, (the magnetic force, levitates, the shaft) will be kept and returned.

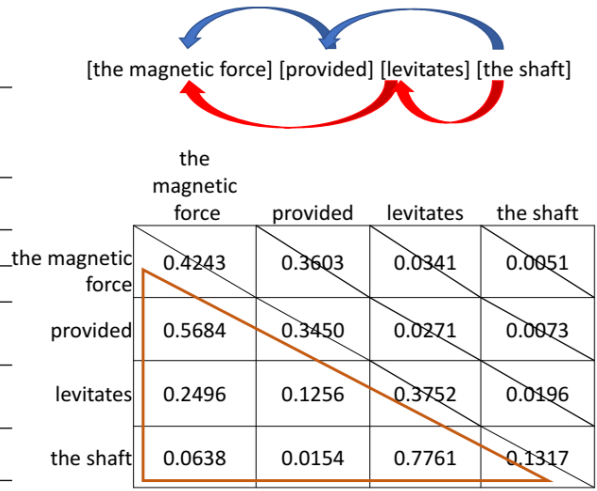

Fig.3.4 The searching process inside the attention graph.

### 3.3.3 Constraints

Patent text can be characterized as comprising long and complex sentences. Various constraints need to be implemented to filter the candidate for higher accuracy.

(1) the threshold of attention score is set to filter the ones with lower reliability. However, each abstract's length varies cause the attention score varies. The threshold is not set as a universal figure but varied on the sentence itself. The median of each abstract' all candidate facts' attention is obtained and used as the threshold.

(2) adv can't be the relationship: adverb naturally can't be candidate relationships, so it is avoided to find more meaningful result.

(3) for a pair of (h, r), multiple tails can be added as a knowledge triple, e.g. (h, r, $t_1$), (h, r, $t_2$).

(4) a h can have one r with highest attention and a tail can only have one r, to avoid the h pair with too much non-sense r. E.g. (h,$r_1$,$t_1$) is permitted, (h, $r_1$, $t_1$), (h,$r_2$,$t_1$) is forbidden.

Fig 3.5 is an example to illustrate the constraints (4). In the sentence "a bearingless hub assembly comprises a rim to receive a tube magnet.", the head "bearingless hub assembly" have two tails "a rim" and "a tube magnet". Within two tails, two verbs are in the between, so totally four candidate facts and there corresponding attention scores are calculated as follows:

|  | a bearingless hub assembly | comprises | a rim | to receive | a tube magnet |
|---|---|---|---|---|---|
| a bearingless hub assembly | 0.5279 | 0.0855 | 0.0131 | 0.0200 | 0.0644 |
| comprises | 0.4030 | 0.2188 | 0.0124 | 0.0078 | 0.0006 |
| a rim | 0.0767 | 0.6746 | 0.1719 | 0.0387 | 0.0108 |
| to receive | 0.0392 | 0.1662 | 0.3131 | 0.4019 | 0.0386 |
| a tube magnet | 0.0076 | 0.0250 | 0.0177 | 0.6368 | 0.3001 |

Fig.3.5 The attention graph to illustrate the constraint (4).

(1) [a bearingless hub assembly] [comprises] [a rim] =0.4030+0.6746= 1.0776

(2) [a bearingless hub assembly] [comprises] [a tube magnet] =0.4030+0.0250=0.4280

(3) [a bearingless hub assembly] [to receive] [a tube magnet] = 0.0382+0.6368=0.6750

(4) [a rim] [to receive] [a tube magnet] = 0.3131+0.6368=0.9499

for 1st, 2nd, and 3rd candidates, head "a bearingless hub assembly" with relationship "comprise" and relationship "receive", the attention score on "comprises" is 1.0776 which is higher than "to receive" 0.6750, so 1st and 2nd candidates are kept. Then for 2nd and 4th candidate, the tail 'a tube magnet' with relationship "to receive" and "comprise", the attention score on 'to receive' is 0.9499 which is higher than 0.4280. Since only one relationship can be kept for one tail, 4th is kept. In this sentence, the knowledge facts (a bearingless hub assembly, comprises, a rim) and (a rim, to receive, a tube magnet) are kept.

## 4. Evaluation

As Table 4.1, the patent-KG extract the knowledge facts from 457,815 patents in Section F from 2016-2021, totally, there are 4,157,377 entities and 10,991,896 edges. The quality of entities and edges are evaluated as follows.

Table 4.1 The size of patent-KG

| Number | Value |
|---|---|
| Number of Patents（F section） | 457,815 |
| Number of Entities | 4,157,377 |
| Number of edges | 10,991,896 |
| Number of phrasal verbs | 317,789 |

## 4.1 Entities

A method similar to that used in Shi's B-link is applied to evaluate the retrieved concepts in patent-KG. A standard list of mechanical engineering related technical terms is built as a benchmark to test whether it's covered in the methods. 3 categories (engines or pumps, engineering in general, lighting and heating) belonging to mechanical engineering are chosen and 13 subcategories, and totally 180 corresponding terms are chosen. Note that the meaning in the terms have some overlapping between each other, the terms in one subcategory can also be classified into another, this list just chose the one of the appropriate subcategories. Table 4.2 listed the categories, the subcategories and some of the terms.

Table 4.2 Standard list of mechanical engineering related technical terms

| Categories | lists |
|---|---|
| **Engines or Pumps** | |
| COMBUSTION ENGINES; HOT-GAS OR COMBUSTION-PRODUCT ENGINE PLANTS | heat engine, combustion chamber, working fluid, pistons, turbine blades, rotor, nozzle., internal combustion engine, four-stroke engine, compression-ignition engine,turbomachinery… |
| MACHINES OR ENGINES FOR LIQUIDS; WIND, SPRING, OR WEIGHT MOTORS; PRODUCING MECHANICAL POWER OR A REACTIVE PROPULSIVE THRUST | pelton wheels, wear-protection couplings, water current turbine, water wheels, Francis turbines, propeller turbines, Kaplan turbines, flywheel, fluid accumulator… |
| **Engineering in general** | |
| FLUID-PRESSURE ACTUATORS; HYDRAULICS OR PNEUMATICS IN GENERAL | pressure intensifier, isobaric pressure exchange, pneumatically operated actuator, hydraulic attachment, hydraulic circuit, positioner, electropneumatic transducer, nozzle-flapper system, pyrotechnic micro-actuator… |
| ENGINEERING ELEMENTS AND UNITS; GENERAL MEASURES FOR PRODUCING AND MAINTAINING EFFECTIVE FUNCTIONING OF MACHINES OR | nails, staples, fastener, rod, anchor, toggle, dowels, bolts, hooks, gear, belts, chains, couplings, cranks, magnetic bearing… |
| INSTALLATIONS; THERMAL INSULATION IN GENERAL | |
| STORING OR DISTRIBUTING GASES OR LIQUIDS | gas filling compartment, gasometers, gas container, gas reservoir , pressure vessels, gas cylinder, gas tank, replaceable cartridge… |
| **Lighting, Heating** | |
| LIGHTING | candle, flash light, illumination, headlight, LED, lamp, Incandescent mantles, pressure vessel… |
| STEAM GENERATION | evaporator, boiler, Inhalator, vaporizer, atomizer, Rankine cycle, working fluid, surface condenser, cooling tower… |
| COMBUSTION APPARATUS; COMBUSTION PROCESSES | Otto cycle, stove, chamber, burner, superheater, reheater… |
| REFRIGERATION OR COOLING; COMBINED HEATING AND REFRIGERATION SYSTEMS; HEAT PUMP SYSTEMS; MANUFACTURE OR STORAGE OF ICE; LIQUEFACTION SOLIDIFICATION OF GASES | freezer, refrigerator, compressor, rectifiers, cryogen, vapor-compression, Stirling cycle, defrost… |
| DRYING | dryer, convection, supercritical drying, dehydration, thermodynamics, moisture, filtration, centrifugation, temperature… |

There are 5 publicly related engineering datasets published before patent-KG, 4 of which are accessible with the API available and can be used as benchmarks for comparison.

$$F = \frac{n}{N} \qquad (1)$$

The results are checked whether are covered in patent-KG and the other 4 methods, to compare the recall rate. The recall rate is calculated as formula (1) where n means the number of covered items in database, while N means the total items. Recall rate is an indication of coverage of the specific field. For example, the wordnet contains 162 of the 180 selected words, therefore the recall rate is 0.9. Table 4.3 shows the total recall rate and the three individual recall rates of the patent-KG and other benchmarks. As can see, the total rate of the patent-KG outperforms the other 3 benchmarks, which means a more engineering specific coverage than other baselines. The patent-KG and the TechNet by Sarica is very close because both focus on patent and have the similar method to extract the terms. Note that patent-KG only choose the patents from 2016 to 2021, with a focus in the 'F' section, while TechNet has a broader time range and broader cover of disciplines. The WordNet(Miller, 1998) and ConceptNet(Speer et al., 2017) performs at a lower recall rate because they focus at a more general level, and single word term contributes a majority in their database.

Specifically, the patent-KG covers more technical terms in the categories of category of 'Engines and pumps', while more specific terms are chosen as examples in this category.

Table 4.3 The comparison between Patent-KG and other baselines

|  | WordNet | ConceptNet | Feng (2017) | Sarica (2019) | Patent-KG |
|---|---|---|---|---|---|
| Total recall rate | 0.46 | 0.56 | 0.63 | 0.79 | **0.82** |
| Engines or Pumps | 0.30 | 0.37 | 0.54 | 0.67 | **0.77** |
| Engineering in general | 0.40 | 0.42 | 0.66 | 0.81 | **0.82** |
| Lighting, heating | 0.68 | 0.88 | 0.70 | **0.88** | 0.86 |

## 4.2 Retrieved relationships

A standard list as Table 4.4 of mechanical engineering related technical relationships is built a benchmark to test whether it's covered in the patent-KG. There are 5 publicly related engineering dataset published before patent-KG. However, some of them don't have semantic relationships (B-link, TechNet), or some of them (Wordnet) is a linguistic English dictionary which mostly provide the taxonomic semantic relationships, such as hypernymy and hyponymy. It's unfair to have a quantifying comparison with a database with different purpose on relationship, so only the result of patent-KG is computed.

Table 4.4 Standard list of mechanical engineering related relationships

| Name | Value |
|---|---|
| relationships | accelerate, add, assemble, block, compute, manufacture, select, prevent, have, made by(with), move, hold, connect to, connect through, include, followed with, reach, pull, lift... |
| recall rate | 0.67 |

### 4.2.1 Qualitative analysis between benchmarks

Sidharth's knowledge graph is in align with the purpose of patent-KG mostly, and it's also based on patent data. However, it didn't have an API for quantifying testing. So a qualitative analysis is conducted based on the relationships in patent-KG and Siddharth's work.

Siddharth applied the rule-based model to extract relationships. The POS tagged VB words as relationships are extracted between nouns. These relationships are classified into hierarchical relationships and non-hierarchical relationships to denote system-subsystem and functional information.

The construction process of patent KG didn't manually list target words or design a syntactic rule to extract the relationship. Instead, the construction process extracts the relationship by applying the attention scores between words. The relationships extracted include not only about system-subsystem information, but also some functional relationships like "for", "increase", "generate", "cast", negation relationship "doesn't", and positional relationship "on", "in".

Besides, the phrasal verbs are recognized as the relationship instead of the verb word alone, thus to provide more contextual information about the relations. For example, "connect to" means bring two things together while "connect through" means establish the connection through something. While these two phrasal verbs mean different, in Siddharth's method, only the single verb word "connect" will be extracted, which can cause an incomplete understanding of the contextual information. For example in the patent of "Automatic high-precision adjustment control system for common electric valve", the knowledge of this patent is extracted as in Fig.4.1 , there are three "connect" related phrasal verbs (connect in, connect between, connect with) and their knowledge triples: (output, connect with, relay), (millisecond-level, connect between, relay) , (flow sensor, connect to, input end)  (frequency conversion module, connect in, output loop). If only "connect" is extract, the positional information in "connect between" and "connect in" is incomplete.

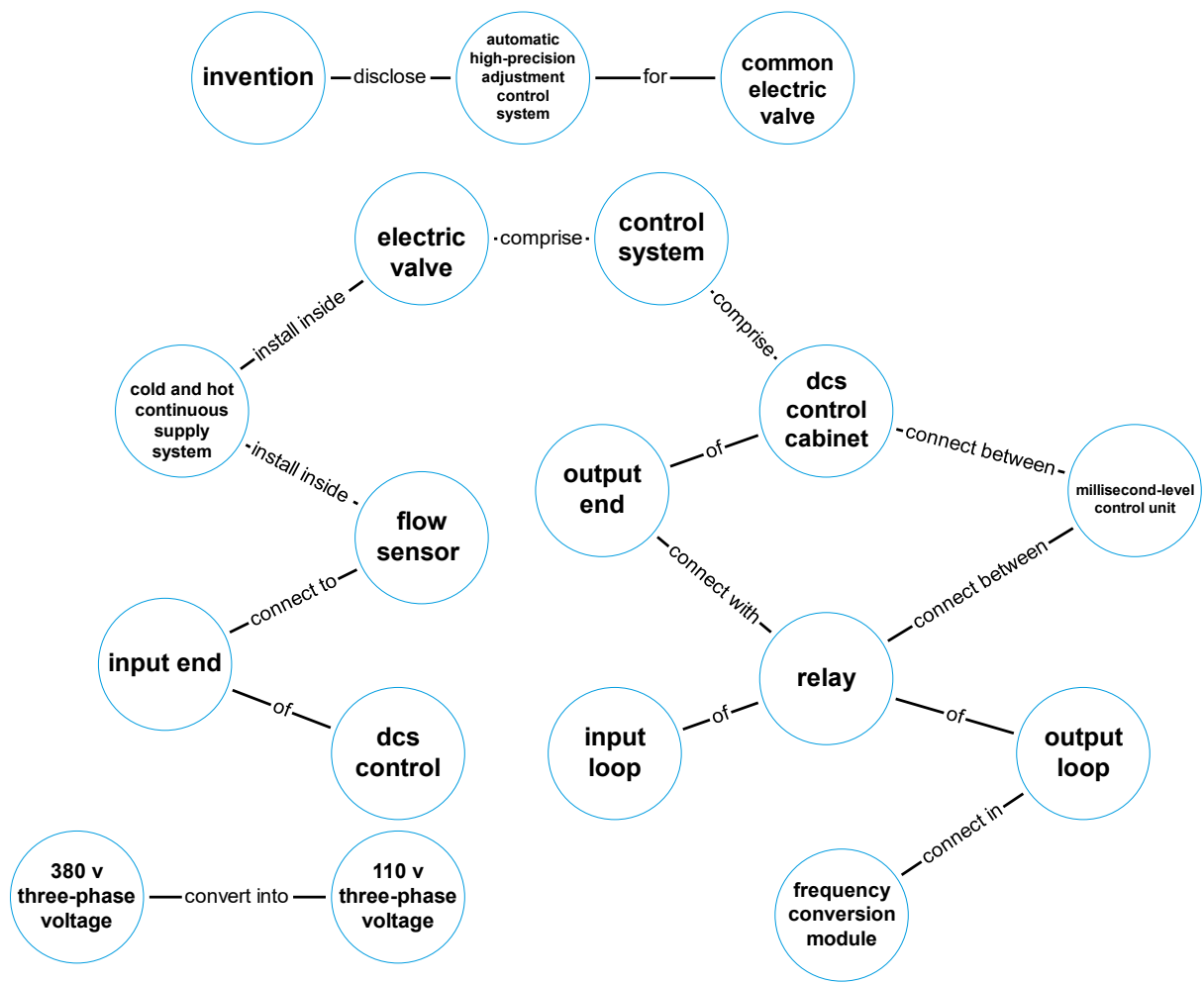

Fig.4.1 The knowledge graph of patent "Automatic high-precision adjustment control system for common electric valve"(Chen Hao, 2019)

Original text in patent is: "The invention discloses an automatic high-precision adjustment control system for a common electric valve, and the automatic high-precision adjustment control system is applied to a cold and hot continuous supply system. The control system comprises an electric valve, a DCS control cabinet and a relay, wherein the electric valve is installed inside the cold and hot continuous supply system, and a temperature sensor, a pressure sensor and a flow sensor are installed inside the cold and hot continuous supply system separately; the temperature sensor, the pressure sensor and the flow sensor are electrically connected to the input end of the DCS control cabinet separately; and the electric valve is connected in an output loop of the relay, an input loop of the relay is connected with the output end of the DCS control cabinet, and a voltage reduction module and a frequency conversion module are electrically connected in the output loop of the relay. A 380 V/50 HZ three-phase voltage input is adopted, 380 V three-phase voltage is converted into 110 V three-phase voltage through the voltage reduction module, and then the frequency is converted to 2 HZ through the frequency conversion module;, and a millisecond-level control unit is electrically connected between the relay and the DCS control cabinet"

By applying to a set of syntactic rules and lexical properties, Siddharth's method theoretically can achieve a higher accuracy rate, which means the relations are the exact and right relations between the head and entities. But because rules are variable and cannot be covered comprehensively, so applying syntactic rules also means a lower recall rate, which means the right relations may not be extracted. Fig 4.2 is a negative example for Siddharth's method. In this example, there are two VERBs between the nouns, while the "provided" actually is a past particle used as an attribute and will cause incorrect knowledge triples. Patent-KG's uses the unsupervised method normally will achieve a higher recall rate, and the accuracy rate is elaborately controlled by setting the threshold and other constraints.

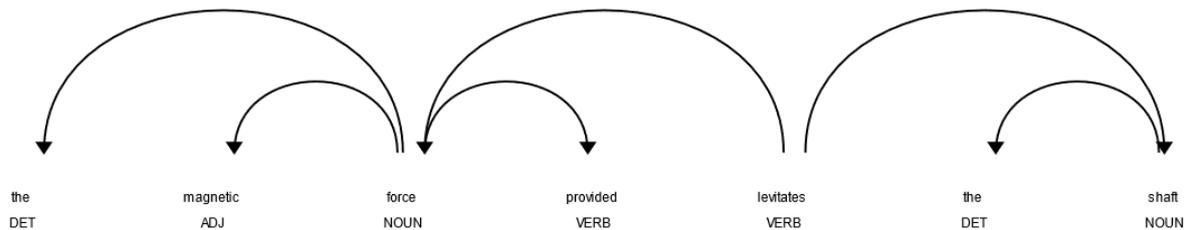

Fig.4.2 The POS tagging example with two VERBs

## 5. Conclusions

A patent can be characterized by its long sentence and complex syntactic structure. With limited labelled data in patents, information extraction tasks can only be done using a traditional rule-based method or an unsupervised processing method. In this paper, we propose an unsupervised method to use the attention mechanism in the language model-BERT, to extract knowledge facts in patents. The quality of the extracted entities and relationships are demonstrated by comparing with other benchmarks. The entities recall rate shows that patent-KG is more engineering specific even with a smaller time scope and disciplines coverage. The relation extraction result suggests that the attention mechanism in language model works for knowledge extraction in patents, without complex rules. Additionally, with qualitative analysis, the attention mechanism extracts phrasal verbs which covers more verb relationships, such as phrasal verbs, positional relation and negation relationships.

This study proposes an unsupervised method to build a knowledge graph for a patent. The unsupervised method can further increase the coverage of an engineering design related knowledge graph, thus facilitating the reuse of knowledge. The proposed method with attention mechanism in this paper is proposed as a starting approach to analyse a patent.

The outcome does not perform well on passive sentences and very long sentences. To achieve better quality and broader coverage of knowledge graph for engineering design, the language models can be fine-tuned over patent text, and the attention mechanism can even be combined with syntactic rules. The knowledge graph is a fundamental step toward knowledge intelligence. In the developing direction of knowledge reuse, enhanced quality directs knowledge reuse from human and computer together as the collaborator, toward automatic knowledge generation with the computer as the creator.